\documentclass{llncs}

\pdfoutput=1

\usepackage{graphicx}
\usepackage{float} 
\usepackage{times}
\usepackage{url}
\usepackage{subfig}
\usepackage{textcomp}
\usepackage{subfig} 
\usepackage{array}
\usepackage{listings}

\begin{document}

\title{A Framework for an Ego-centered and Time-aware Visualization of Relations in Arbitrary Data Repositories}

\author{Florian Reitz}

\institute{Dept. of Databases and Information Systems\\ University of Trier, Germany \\  \email{reitzf@uni-trier.de}
}
\maketitle 

\begin{abstract}
Understanding constellations in large data collections has become a common task. One obstacle a user has to overcome is the internal complexity of these repositories. For example, extracting connected data from a normalized relational database requires knowledge of the table structure which might not be available for the casual user. In this paper we present a visualization framework which presents the collection as a set of entities and relations (on the data level). Using rating functions, we divide large relation networks into small graphs which resemble ego-centered networks. These graphs are connected so the user can browse from one to another. To further assist the user, we present two views which embed information on the evolution of the relations into the graphs. Each view emphasizes another aspect of temporal  development. The framework can be adapted to any repository by a flexible data interface and a graph configuration file. We present some first web-based applications including a visualization of the DBLP data set. We use the DBLP visualization to evaluate our approach.
\end{abstract} 

\section{Introduction}
\label{sec:introduction}

The amount of data we create and store increases every day. The result is a growing number of large and possibly complex repositories. Much effort has been invested in defining potent query languages like SQL and XQuery and there are many fast and reliable algorithms to apply them. However, these techniques do little to hide the internal structure of the collection. For example, a highly normalized relational database might require several table joins to gather all information which belongs to one entry. While an expert can use information on the internal structure to speed up queries, a casual novice might easily be overstrained. This is a problem because the user wastes valuable time needed for the actual task on understanding the data organization.

A common way to hide internal complexity is to present the collection as a set of entities which are connected by various types of relations. Consider the DLBP bibliographic data set\footnote{dblp.uni-trier.de}. It contains meta data for 1.3 million publications. \textit{Authors} are one type of entity that can be found in the collection. They are connected by the \textit{coauthor} relation if they cooperated on at least one publication. There are several approaches to use this model in applications. Some prefer a textual representation \cite{BastCSW07} \cite{KuhnW08} while others provide a more complex visualization\cite{LeeCRB05} \cite{WeberRWLK06}. Relations also provide a context for a given entity and thereby support exploring the data set. 

\begin{figure}
 \centering  
  \includegraphics[width=8cm]{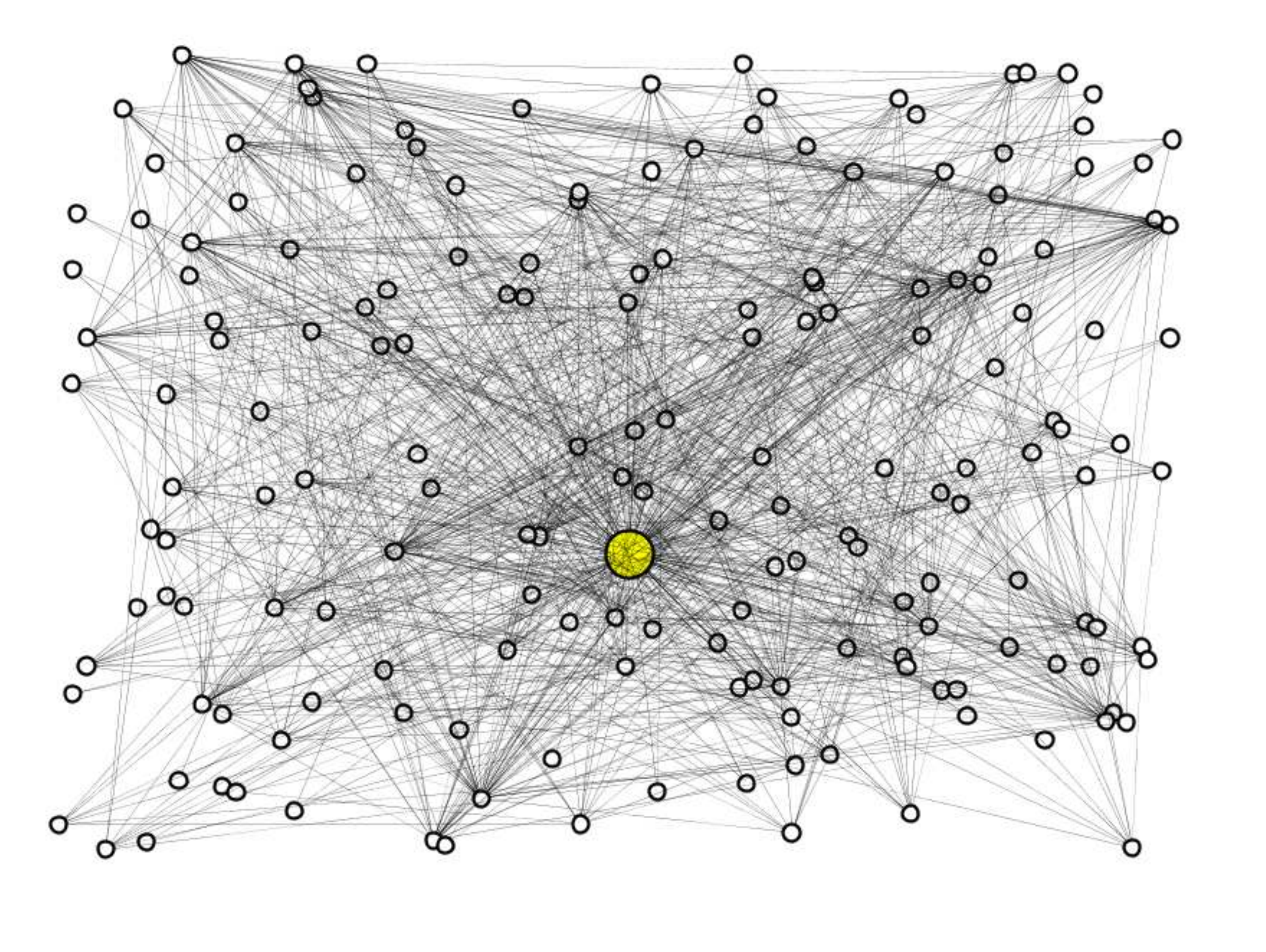}  \label{img:introA}  
  \caption{\textit{Adam's} complete neighborhood in the DBLP coauthor graph. \textit{Adam's} node is painted in yellow}
  \label{img:intro}
\end{figure}

The networks which are defined by relations in real-world data sets can be very large and dense. The DBLP coauthor network consists of 750,000 authors and 2.5 million relations. Suppose we want to examine the publication behavior of author \textit{Adam}. Even finding \textit{Adam} in this network is cumbersome. Figure \ref{img:intro} shows a node-link drawing of \textit{Adam's} direct neighborhood in the coauthor network. It still contains 178 nodes and 
1154 edges (177 \textit{Adam} - coauthor + 977 coauthor - coauthor). The drawing hides the fact that some relations are stronger than others. For example, \textit{Adam} has cooperated 25 times with \textit{Bob}, but only 1 time with Jack, so the tie to \textit{Bob} is much stronger. It also does not reveal 
at which time the relations were formed and how they have evolved over time.

In this paper we show how relation networks can be divided into small connected graphs. In Section \ref{sec:relationgraphs} we describe how filtering and rating can reduce the neighborhood graphs. We obtain simple graph drawings which resemble ego-centered networks \cite{Freeman82}. These drawings can be enriched with information on how a relation has evolved. In Section \ref{sec:timeviews} we present two views which cover different aspects of evolution. One emphasizes \textit{when} a relation was influenced while the other focuses on \textit{how strong} this influence was. A single graph is of little use if it is not linked with others. In Section \ref{sec:framework} we show how the graphs are embedded into a framework which allows browsing the graphs. In Section \ref{sec:examples} we present examples of visualized relations in DBLP and the German language Wikipedia. We conclude this paper with a two-part evaluation of our approach.

\section{Relation graphs}
\label{sec:relationgraphs}
A common approach to visualize graphs is the node-link diagram. Entities are represented by nodes. If two entities are in relation, their nodes are connected by an edge. In most real-world data sets there are a number of different relations. The networks defined by these relations can become very large and do not allow comprehensible node-link drawings. To reduce their size, we generate a set of ego-centered drawings.

\subsection{Inverted Ego Graphs}
\label{sec:egographs}

\begin{figure}
 \centering  
  \subfloat[ego centered network]{\includegraphics[height=4.2cm]{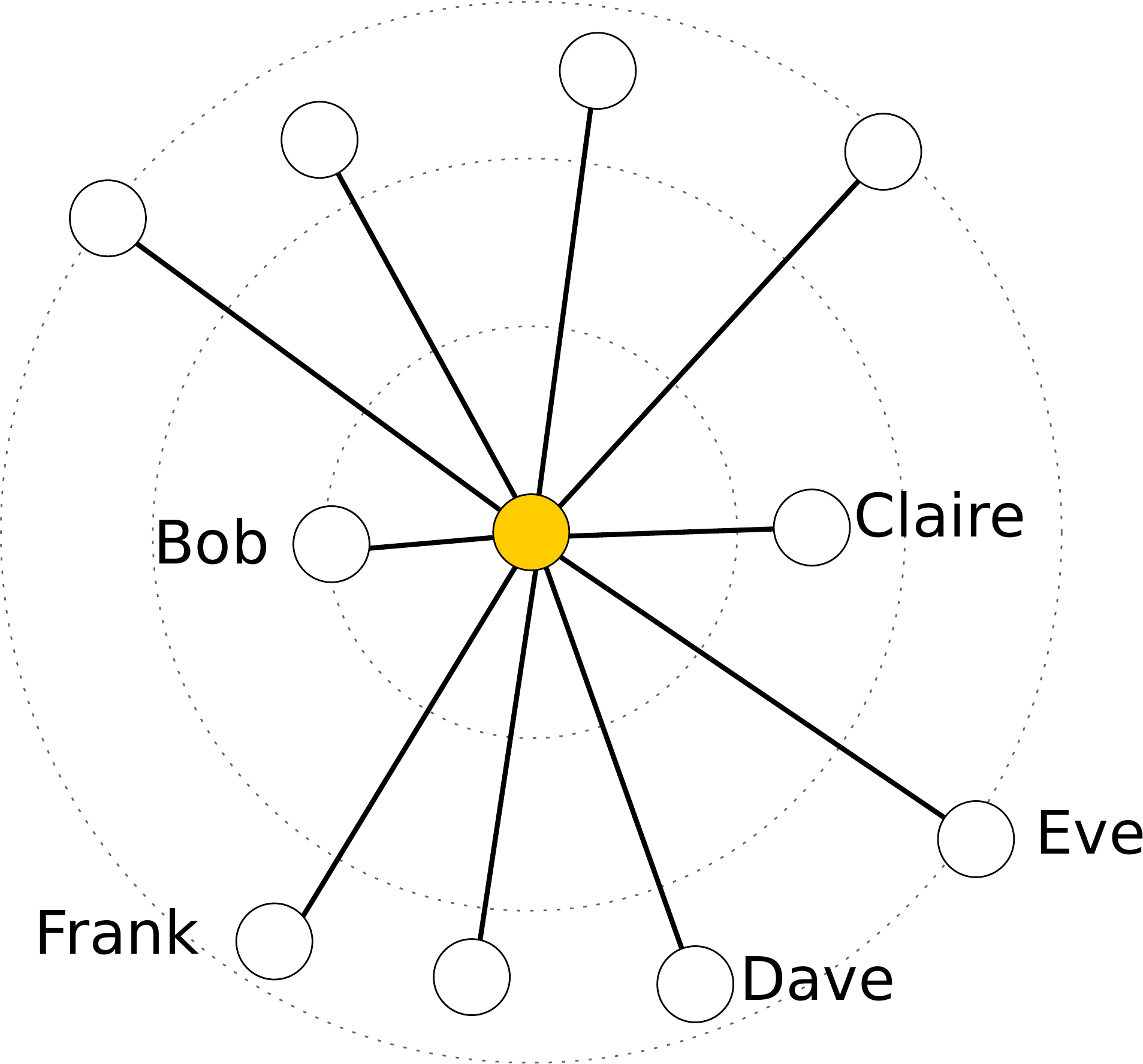}  \label{img:egoA}} \qquad
  \subfloat[inverted ego graph]{\includegraphics[height=4.2cm]{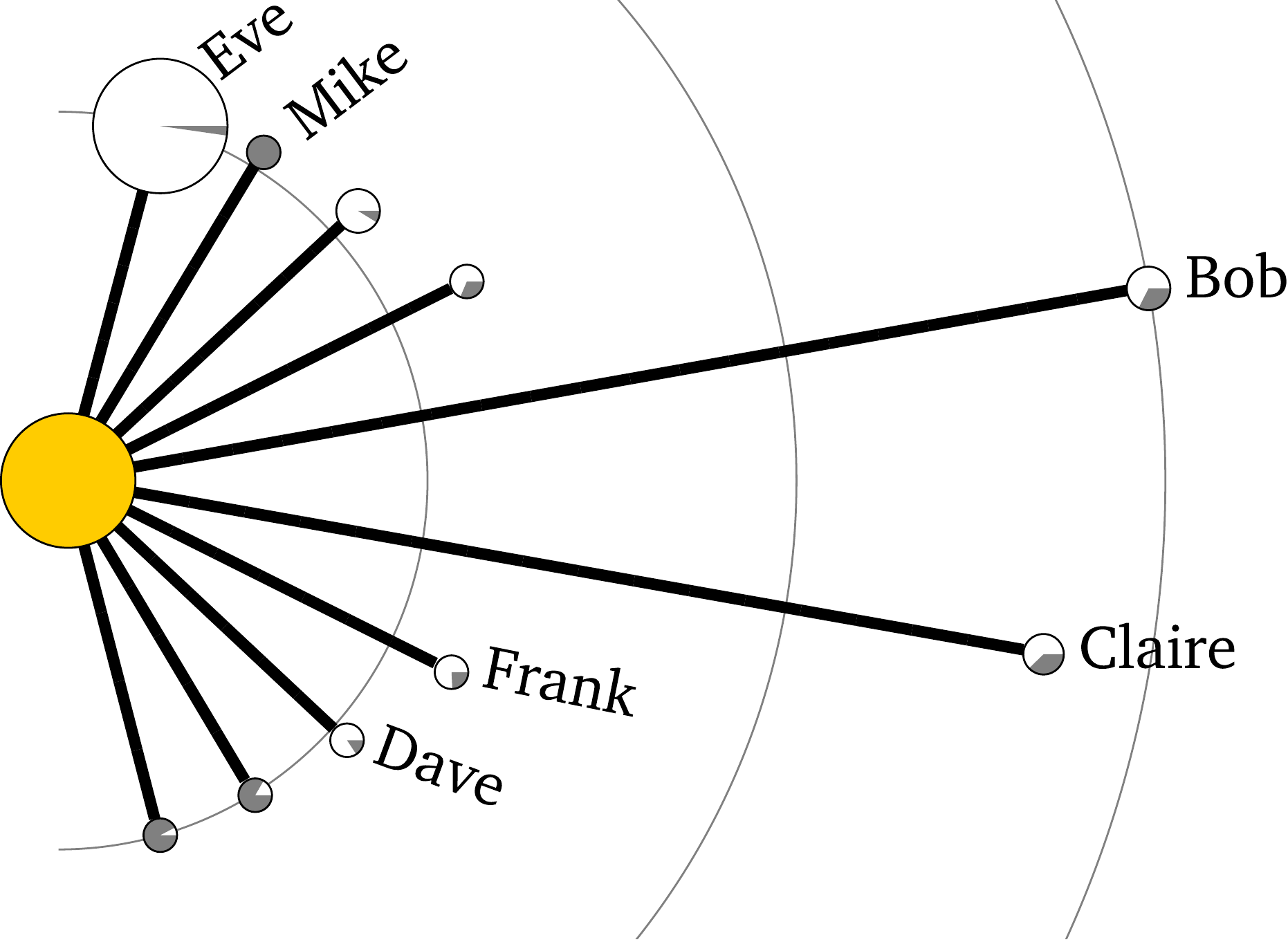}  \label{img:egoB}}
  \caption{\textit{Adam's} coauthor network as ego graphs. \textit{Adam's} node is painted in yellow.}
  \label{img:ego}
\end{figure}

In a first step we split the relation networks into small graphs. Each graph represents the direct proximity of one entity. We call this entity the \textit{ego} of the drawing. In the coauthor example, we obtain 750,000 graphs, one for each author. Figure \ref{img:intro} shows that this approach will be virtually useless if the drawing violates several aesthetic criteria which have been identified for this type of drawing \cite{BennettRSG07} \cite{Purchase2002501}. Above all, the number of edge crossings is problematic here. To improve the comprehensibility of the split graphs we only include the most relevant neighbors. For most relations there is a straightforward definition of relevance. In the coauthor example, we can use the number of joint publications. If we put the node which represents ego in the center of the drawing and place the selected neighbors (the \textit{alters}) in a distance according to their relevance, we will obtain an ego-centered network \cite{Freeman82}. We do not draw edges between alters to avoid edge crossings. Figure \ref{img:egoA} shows \textit{Adam's} ego-centered network. We have included the ten most relevant alters. \textit{Bob} is placed closest because the relation with him is the strongest. \textit{Eve} is the least relevant of the included authors.

In Section \ref{sec:timeviews} we will use the edges to display information on how the relation has evolved over time. In many cases, the most relevant alter  is of special interest and we can expect that the development of this relation is more complex than others. However, the edge which represents this relation is the shortest in the drawing and thereby provides the least space for additional information. To increase the available space, we introduce the \textbf{inverted ego graph}. We place the most relevant alter at maximum distance to the ego while less important alters are positioned closer. To further increase the edge length, we place the ego and the most relevant alter at opposite sides of the drawing. This contradicts the \textit{close related entities are placed close to each other} metaphor, but it provides sufficient space. Figure \ref{img:egoB} shows \textit{Adam's} inverted ego graph.

\subsection{Drawing Details} 
\label{sec:drawingdetails}

\begin{table}
\centering
\begin{tabular}{|m{1.2cm}|m{2.9cm}| m{3cm}|}
\hline
\textbf{type} & \textbf{parameters} & \textbf{examples} \\
\hline
none & - &  \includegraphics[width=0.7cm]{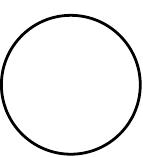}  \includegraphics[width=0.7cm]{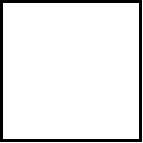} \includegraphics[width=0.7cm]{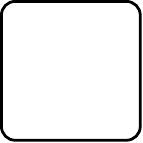}  \includegraphics[width=0.7cm]{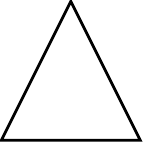} \\
\hline
solid & \textit{one color} &  \includegraphics[width=0.7cm]{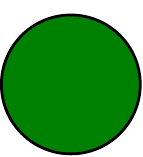}  \includegraphics[width=0.7cm]{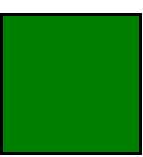}  \includegraphics[width=0.7cm]{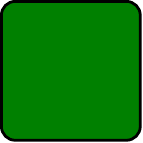}  \includegraphics[width=0.7cm]{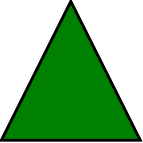}\\
\hline
fraction & $d \in [0,1]$, \textit{one color}  &  \includegraphics[width=0.7cm]{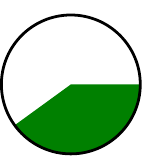}  \includegraphics[width=0.7cm]{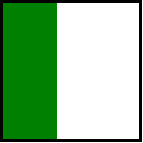}  \includegraphics[width=0.7cm]{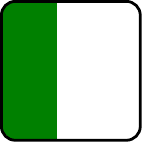} \includegraphics[width=0.7cm]{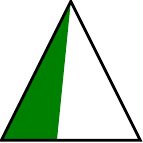}\\
\hline
pie & Map (color $\rightarrow$ double) &  \includegraphics[width=0.7cm]{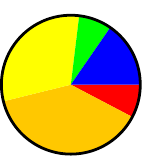}  \includegraphics[width=0.7cm]{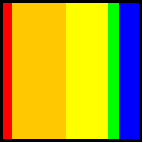}  \includegraphics[width=0.7cm]{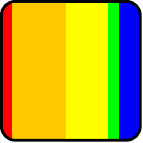}  \includegraphics[width=0.7cm]{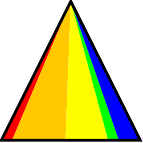}\\
\hline
timecolor & list of colors &  \includegraphics[width=0.7cm]{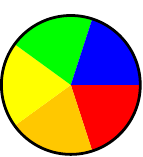}  \includegraphics[width=0.7cm]{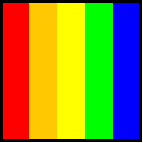}  \includegraphics[width=0.7cm]{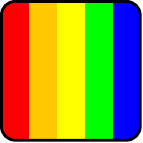}  \includegraphics[width=0.7cm]{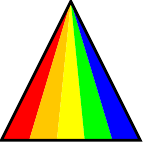}\\ 
\hline
presence & list of Booleans, \textit{one color} &  \includegraphics[width=0.7cm]{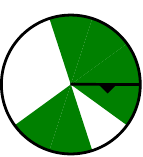}  \includegraphics[width=0.7cm]{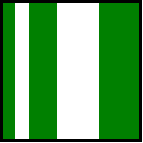} \includegraphics[width=0.7cm]{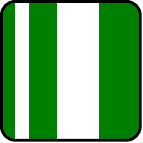}  \includegraphics[width=0.7cm]{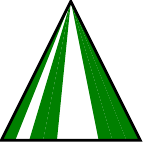}\\
\hline
\end{tabular}
\caption{The available node fillings. Parameters in italics are optional.}
\label{tab:details}
\end{table} 

Ego-centered networks originate from the field of qualitative network analysis. Aside from the relevance value, these drawings often contain additional information which might be important for a user who interprets them. We can add information by modifying size, shape and filling of the nodes. For example, in Figure \ref{img:egoB} we use the size of the alters to show the total number of publications. Obviously, \textit{Eve} is by far the most active author. However, she has a low relevance because only a small part of her publications were joint work with \textit{Adam}. We use the node filling to show this fraction. \textit{Mike} for example published all work with \textit{Adam}. Together with the low number of publications, this suggests that he is a PhD student rather than an established researcher. Both ego and alters represent persons so we use the same node shape to draw both.

Table \ref{tab:details} shows the six types of node fillings and the parameters they require. \texttt{none} leaves the nodes empty while \texttt{solid} fills them with a specified color. \texttt{fraction} requires a value $d \in [0,1]$ which defines how much of the node area is filled. \texttt{pie} is a generalization of fraction where multiple color segments can be used. In the \texttt{time color} filling, all segments have the same size and the whole area is covered. The  \texttt{presence} filling divides the area into equal-sized parts. A list of Boolean values defines which parts are filled and which are left white. In Section \ref{sec:examples} we will show examples of all types.

\section{Time Views}
\label{sec:timeviews}
 
\begin{figure}
 \centering  
  \includegraphics[height=5.5cm]{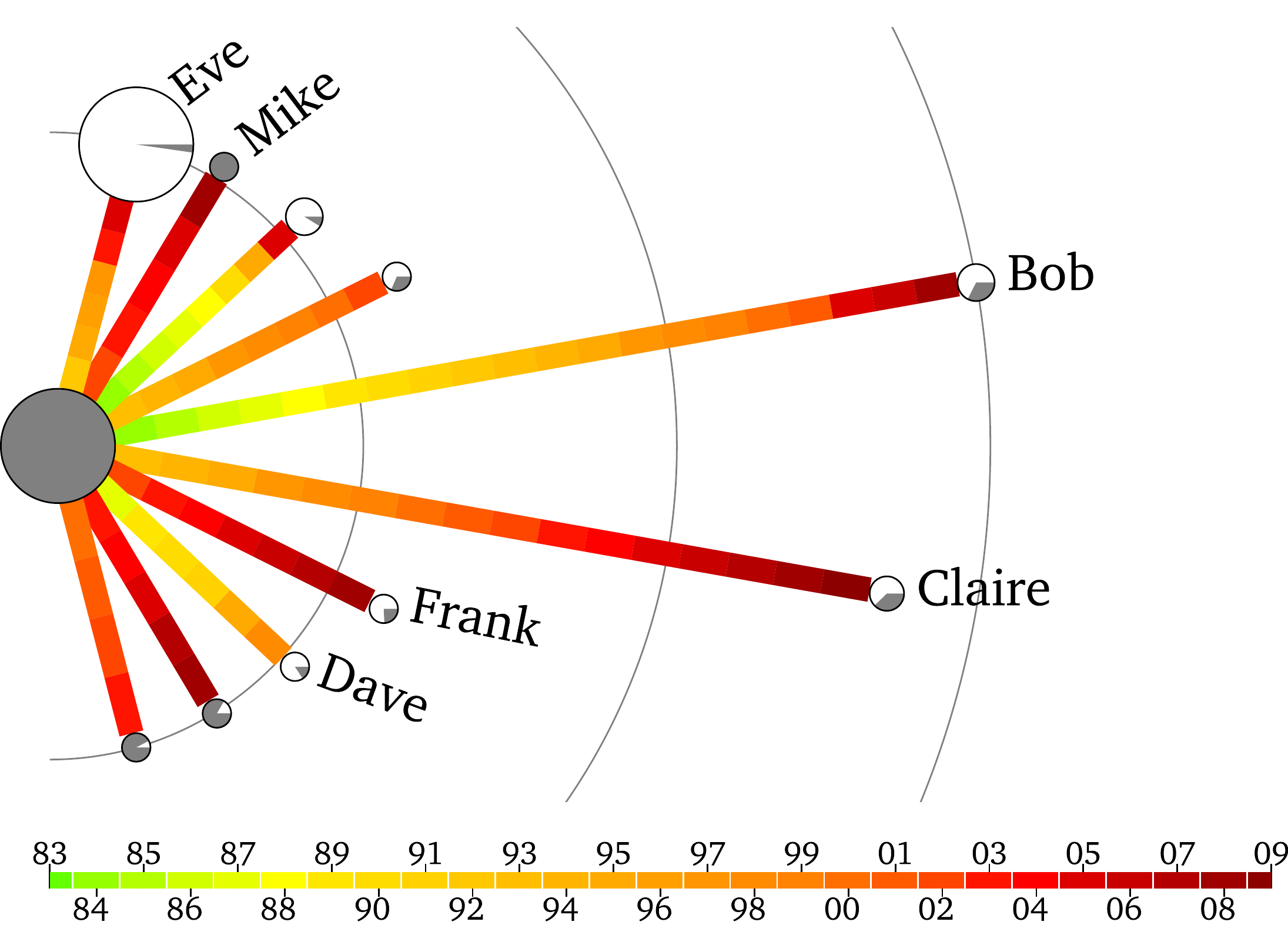}  \label{img:introC} 
  \caption{The time-color view for \textit{Adam's} coauthors}
  \label{img:time}
\end{figure}

Not all relations have evolved in the same way and a user might be interested to see these differences. For example, if we look for \textit{Adam's} long-term partners, we want to discriminate them from coauthors with a short but intense cooperation. First, we divide the time frame, i.e., the interval between the oldest and the youngest time stamp in the data set into equal-sized periods. The data set must contain sufficient information to determine how strong a relation was at each period. In the DBLP example, the time stamps represent the year of publication. The oldest paper is from 1936, the newest from 2010. We use 75 periods, each covering a single year. The strength of a relation in a single period is the number of joint papers in that year. This information is available in the data set. There are periods in which the relation becomes stronger and periods in which it remains unchanged. We do not consider relations which become less intense over time.

We use the edge between ego and alter to show the development in different periods. There are two aspects we must consider: \textit{when} was the relation influenced and \textit{how strong} was this influence. For each aspect we have implemented a view, i.e., a way to modify the edge drawings. The positions of the nodes are the same for both views so we can easily switch between the two to see both aspects.

\begin{figure}
 \centering  
  \subfloat[time-color view]{\includegraphics[width=13cm]{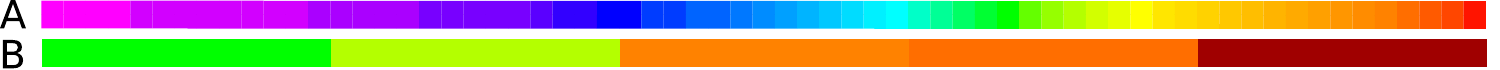}  \label{img:edge_color}} \\
  \subfloat[intensity view]{\includegraphics[width=13cm]{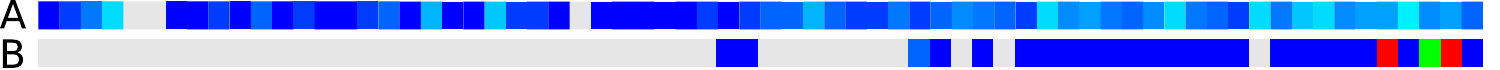}  \label{img:edge_intensity}}
  \caption{Two edges from time-color view and intensity view respectively with different development}
  \label{img:edges}
\end{figure}

\subsection{Time-Color View}
\label{sec:timecolor}

\begin{figure}
 \centering  
  \subfloat[\textit{Adam}. 2005 is highlighted]{\includegraphics[height=5.5cm]{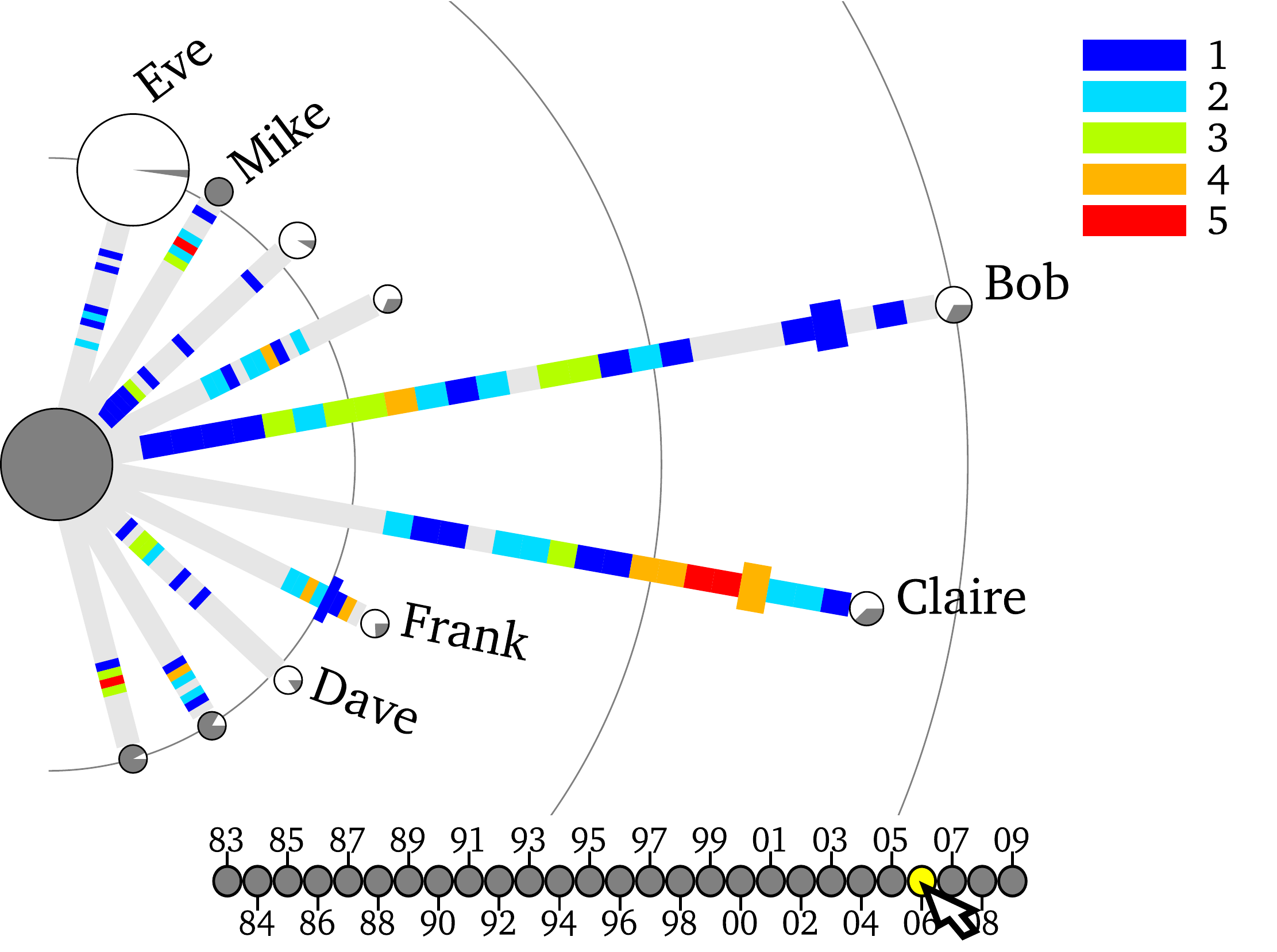} \label{img:intensityA} } \\
  \subfloat[\textit{Nicole}. Values 15 - 16 are highlighted]{\includegraphics[height=5.5cm]{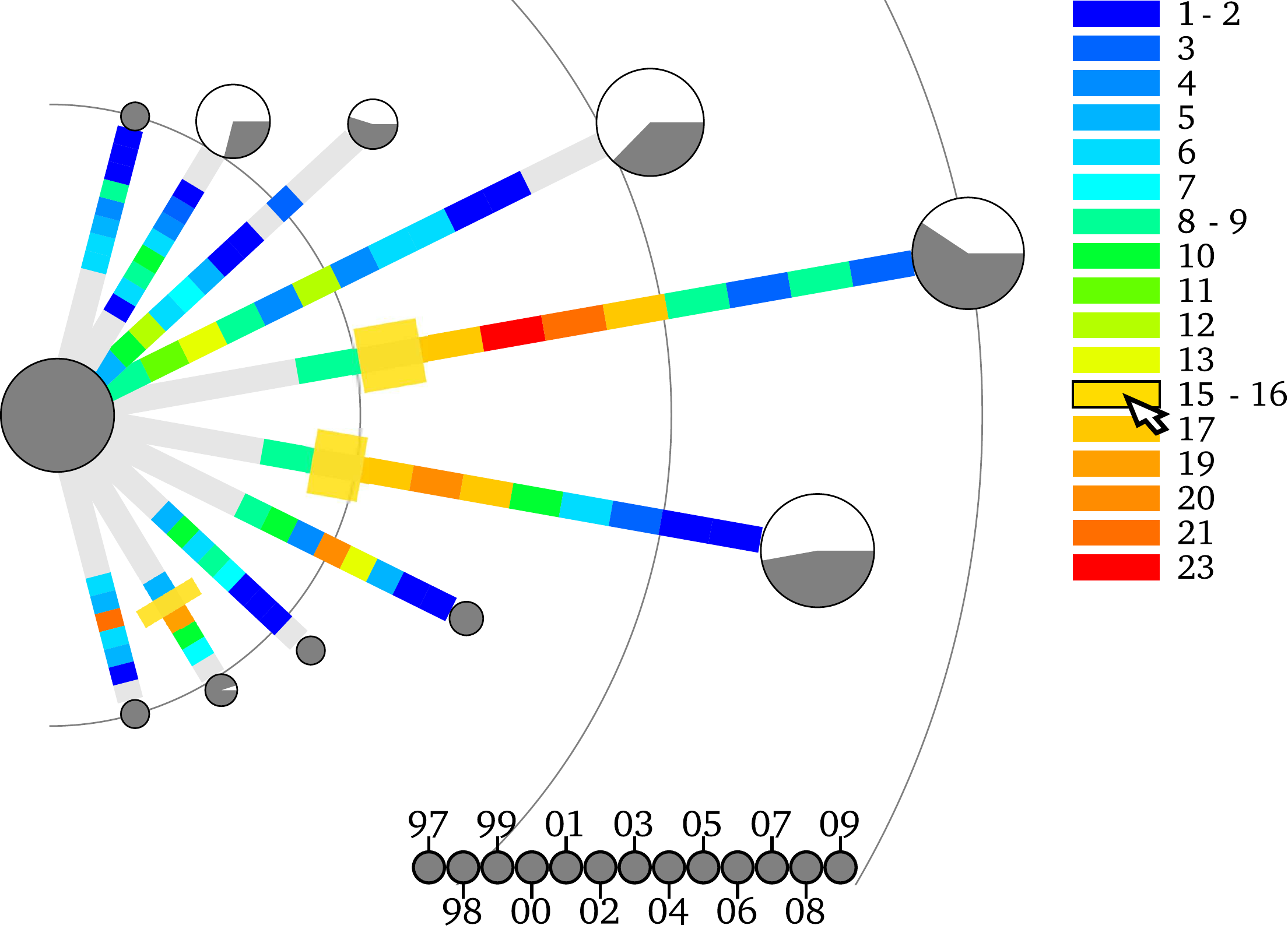} \label{img:intensityB}}
  \caption{The inverted coauthor ego network of \textit{Adam} and \textit{Nicole} in the intensity view}
  \label{img:intensity}
\end{figure}

The time-color view emphasizes the moment in which a relation was influenced. We assign a color to each period which expresses its position in the time frame. There is no linear order of colors but most people accept the purple-blue-green-yellow-orange-red sequence for this purpose \cite{Ware88}. We assign purple and blue tones to periods from the beginning of the time frame. Red tones indicate recent periods. For each period which affected the relation, regardless how strong, we add a colored segment to the respective edge. The segment color matches the represented period. Periods with no influence have no representation. The segments are ordered by time, the oldest close to the ego and the newest close to the alter. Figure \ref{img:time} shows \textit{Adam's} coauthors in the time-color view. At the bottom, there is a color bar which shows the time-color mapping. Green and yellow segments refer to the 1980s while reddish segments represent the years after 2000. We can see that the relation between \textit{Adam} and \textit{Bob} has evolved over a long time while the cooperation with \textit{Claire} started later. The cooperation with \textit{Dave} ended in the late 1990s. 

The segment size depends on the number of periods which are relevant for the respective relation and the available edge length. Figure \ref{img:edge_color}
shows two edges with different content. While edge A covers many consecutive periods, B shows two distinct phases of development marked by an abrupt color change. In a first version of this view, we modified a segment length based on the strength of influence. The segments of important periods became longer compared to those of less relevant periods. Early user feedback (see Section \ref{sec:fieldstudy}) showed that this additional distortion confused the viewers. We gave up on this feature in favor of the intensity view.

The number of periods is limited by the number of colors a human can discriminate. In an ideal environment, we can distinguish more than one million colors\cite{KaiserB96} but the user feedback showed that even the 75 DBLP periods can become problematic. Especially, matching segment colors with the bottom bar was reported to be difficult. We use linking and brushing to compensate. Whenever the mouse cursor moves over a segment, all segments of the same color are painted in double stroke. The bottom bar shows only those segments which are relevant for at least one relation. 

\subsection{Intensity View}
\label{sec:intensity}

The intensity view emphasizes how strong a relation was influenced during a period. We use colors to show the strength of development. Periods with little influence are painted in blue while more important ones are presented in reddish tones. Unlike the time-color view, each edge contains segments for all periods. Like before we leave out periods which are not relevant for any relation. If a period had no influence on a relation, the segment is painted in white so it appears as a gap. Because the number of segments is the same for all edges, one level of distortion is eliminated. Now we can use the position of the segments to guess the period. Cleveland and McGill \cite{Cleveland1986491} showed that humans can perceive positions much better than colors. This compensates for the smaller segments. Figure \ref{img:edge_intensity} shows two edges from the intensity view. While edge A shows a homogeneous but weak development with only three gaps, edge B represents a younger relation with a strong recent development.

Figure \ref{img:intensityA} shows \textit{Adam's} coauthor graph in the intensity view. The strength of development in a period is equivalent to the number of joint publications in this year. Although \textit{Bob} and \textit{Claire} have similar importance values, we can clearly see the differences in development. It is also visible that the last cooperation between \textit{Adam} and \textit{Dave} happened some time ago. An additional legend at the right side of the graph maps colors and values. In Figure \ref{img:intensityA} there are only five different development values. In Figure \ref{img:intensityA} all segments of the year 2005 are highlighted by brushing over the bottom bar. In Figure \ref{img:intensityB} we see the coauthors of \textit{Nicole} who published 23 papers in cooperation with a single author in a single year. If the domain becomes too large, the framework maps intervals to a color instead of single values. In this drawing, all segments with 15 or 16 joint papers are highlighted. 

\section{Framework Components}
\label{sec:framework}
The framework consists of three major components, the user interface, the graph generator and the interface with the underlying data.

\begin{figure} 
\centering
\includegraphics[width=12cm]{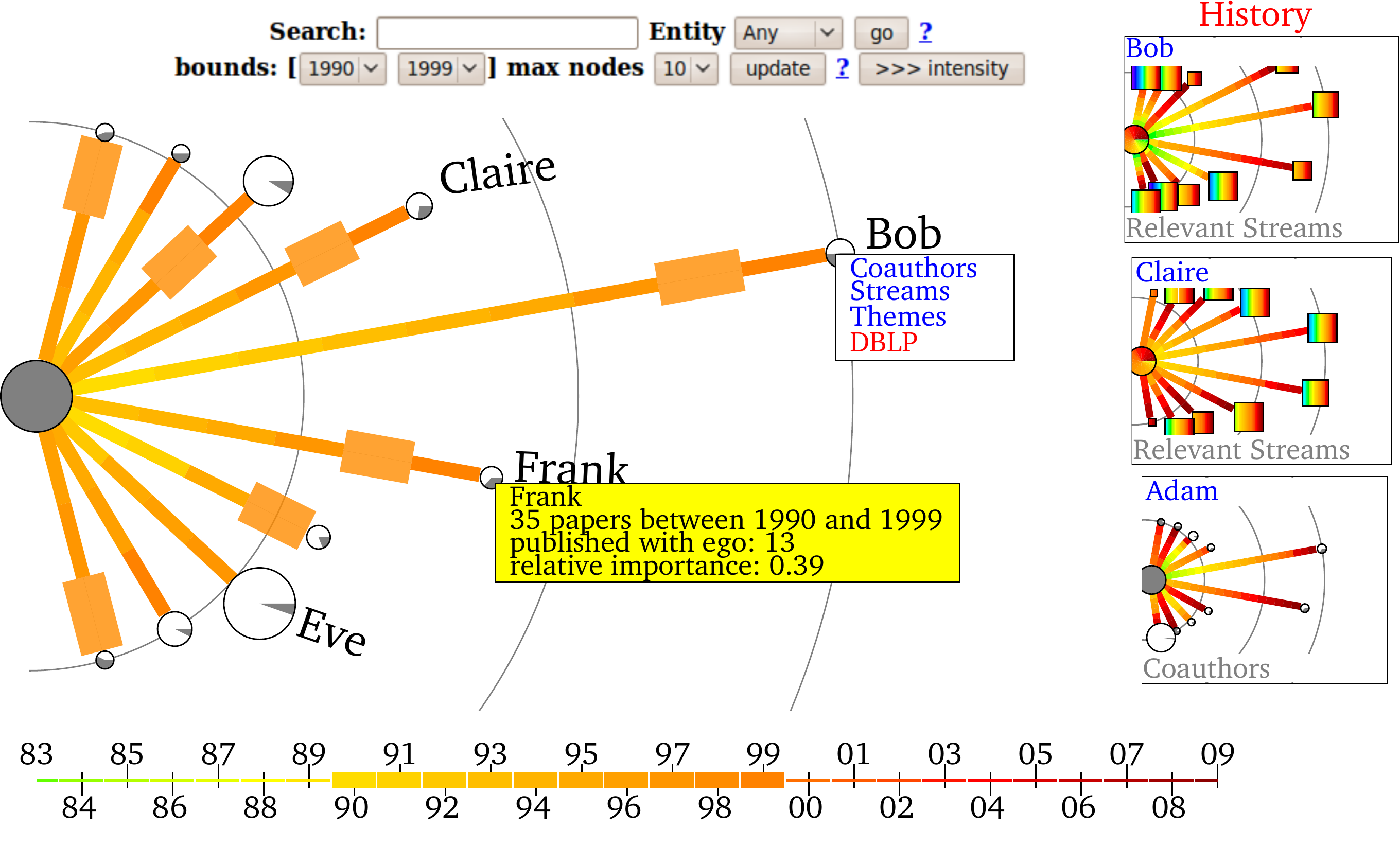}
\caption{The user interface showing an example of \textit{Adam's} coauthor graph in the time-color view.}
\label{img:interface}
\end{figure}

The front end is composed of the graph drawing and some additional elements which assist the user. We use the SVG (Scalable Vector Graphics) format for the drawings. Vector graphics are smaller then pixel graphics and can be zoomed without loss of quality. There are a number of frameworks which allow SVG rendering in applications and many web browsers support a sufficient part of the standard. Figure \ref{img:interface} shows an example of a web-based front end displayed by the Firefox 3.0 browser. The component positions are a suggestion and can be changed if needed.

The most important front end function is to link the graphs. We can click on a node and a new graph is created where this node is the ego. In Figure \ref{img:interface} there is more than one possible relation for a person type ego. A connection menu (Bob) lists the available options. In the example, there is also an external link to \textit{Bob's} DBLP author page. In theory, SVG graphics can provide smooth changeovers between the graphs where nodes move to their new positions and new ones are faded in. However, only few browsers support SVG animation. We expect this to improve in the future.

There are two other ways to switch to a new graph. The head menu provides a search field which can be used to find a new ego if the node is not contained in the current graph. The menu also contains control elements for the time lens feature. The time lens allows the user to define which periods should be considered for displaying the graphs and computing the relevance function. In Figure \ref{img:interface} we see \textit{Adam} as if he only published in the 1990s. The bottom bar shows which periods are relevant here. If we compare the relevance values with those in Figure \ref{img:time} we clearly see some differences. The head menu also contains a control to modify the maximum number of alters. Often, it is useful to go back to a previous graph. The history bar shows thumbnails of up to four old requests. We can click the thumbs to enlarge the drawings again. In some situations it is useful to have textual representation of data. All nodes have a tooltip menu which appears when the cursor moves over it. In the next section, we will show how they are defined. Like all visual effects, including those we discussed in Sections \ref{sec:timecolor} and \ref{sec:intensity}, tooltips are created by JavaScript functions which are embedded into the SVG files. No additional command processing is necessary.

There is no limit to the complexity of the underlying data source. In most cases we need sophisticated information extraction strategies and hand tuned queries to achieve acceptable performance. The data which is necessary to draw an entity or a relation might be scattered over the whole data set. Only an expert can provide a fast interface with these repositories by implementing a given Java interface. An important part of this step is to actually define the entities and relations. The expert can utilize caching strategies and pre calculations if necessary and enforce privacy policies for example by making entities anonymous. Other framework components request data from the interface by giving a pre-defined key and a data type. A similar technique is used to define the rating function for the relations.

\section{Example Applications}
\label{sec:examples}
To demonstrate the framework, we deployed two applications on data sets with different characteristics: DBLP and the German language Wikipedia.

\subsection{DBLP}
\label{sec:dblp}
The coauthor graph drawings we used in this paper are taken from this application. In this section we give some background information and additional examples for drawing details. This visualization is available online\footnote{http://dblpvis.uni-trier.de/} and generated some user feedback. The DBLP data set is available as an XML file which lists all records. The size and the structure of this file make efficient queries impossible so we import the document into a relational database. During this step, we pre calculate some frequently used values. We take additional information on journals and conferences from html pages on the DBLP web server.

\noindent \textbf{Entities: } We extract three types of entities (number on December 8, 2009): \textit{person} (760,277), \textit{word} (36,150) and \textit{stream} (3480). Stream is the term DBLP uses to refer to conferences and journals. Each author with at least one publication listed in DBLP is represented by a \textit{person} entity. DBLP uses person names as identifiers. If there are two homonymous authors, a numeric suffix is appended to the name. If a person uses different names the data set contains entries to match them. The search engine is customized to consider this additional data as well. 
The \textit{word} entities are derived from the publication titles. We remove stop words and invert inflections.

\begin{figure}
 \centering  
  \includegraphics[height=7cm]{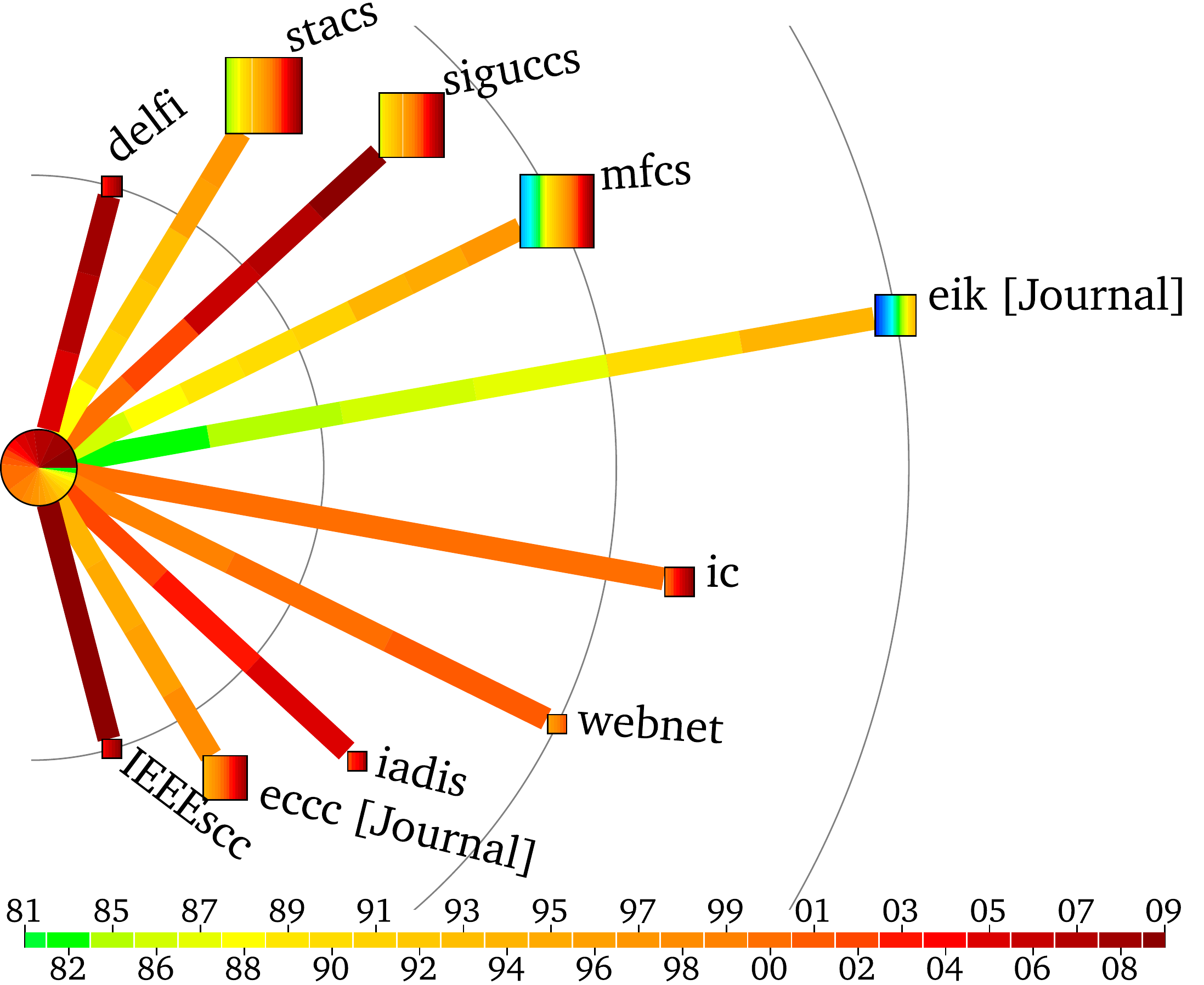}
  \caption{\textit{Petra's} relations to conferences and journals in the time-color view}
  \label{img:personstream}
\end{figure}

\noindent \textbf{Relations: } Figure \ref{img:personstream} shows the relations between \textit{Petra} and the streams which accepted her papers. The relevance for this relation is defined by the number of accepted papers. The edge coloration shows that \textit{Petra} stopped publishing at \textit{eik}, \textit{mfcs} and \textit{stacs} in the 1990s. To better understand the reasons for this, we use a time-color filling for the stream nodes. For each year a conference or a journal was active (held a venue or published an issue) we add a segment in the respective color. If a stream is old some colors will not be shown in the bottom bar. Missing red segments show that \textit{eik} was not continued after 1994 which explains why there are no further publications. However, \textit{mfcs} and \textit{stacs} were continued after \textit{Petra's} last publications so there must be other reasons which the data set does not reveal. With the exception of \textit{SIGUCCS} all stream nodes with recent publications are small. This means they do not have many years of activity and are rather young. The pie filling of the ego node shows when \textit{Petra} was active. The more papers she published in a year the larger the respective segment. 

\begin{figure}
 \centering   
  \includegraphics[height=7cm]{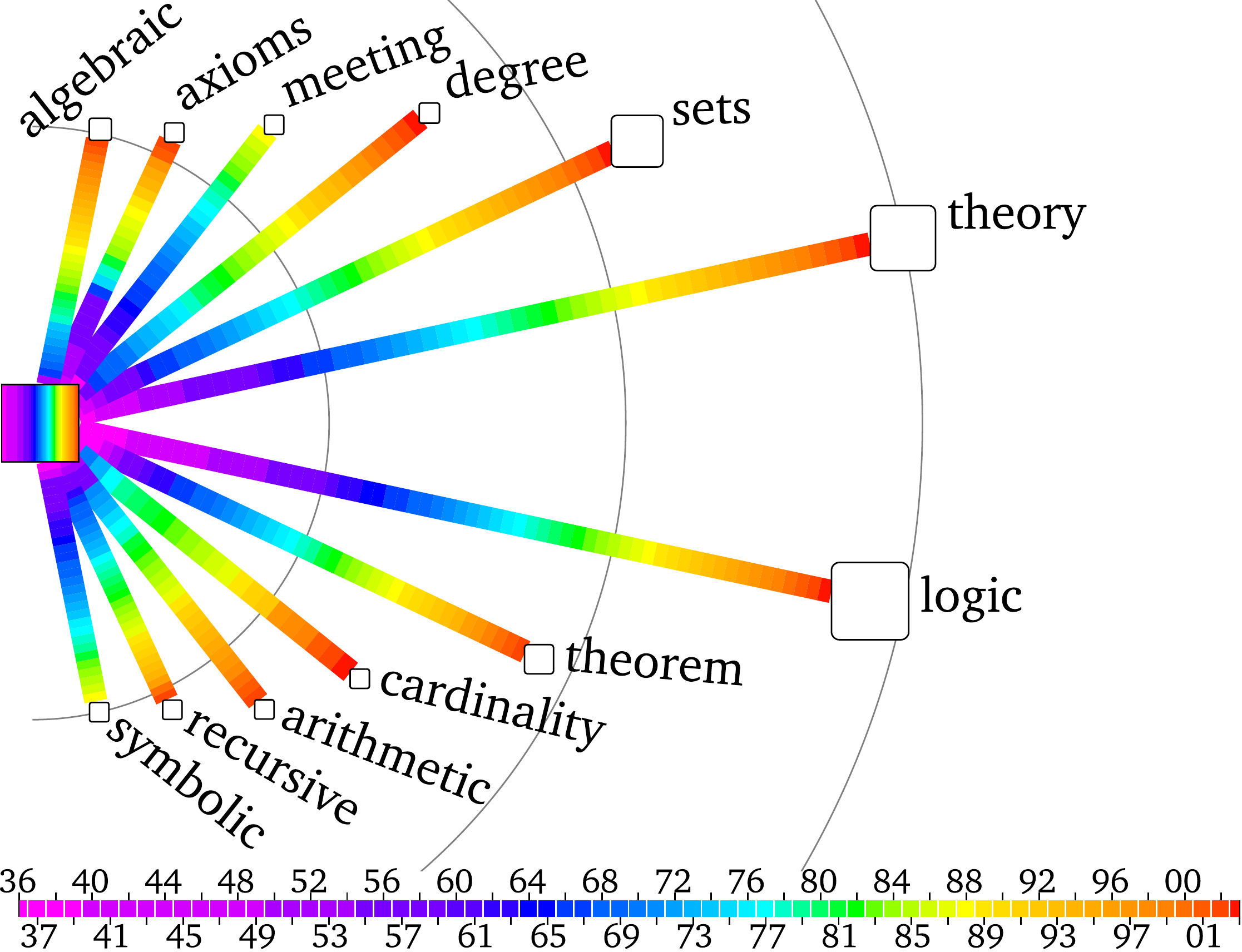} 
  \caption{Related themes of \textit{JSYML} in the time-color view} \label{img:streamword}  
\end{figure}

\begin{figure}
 \centering   
  \includegraphics[height=7cm]{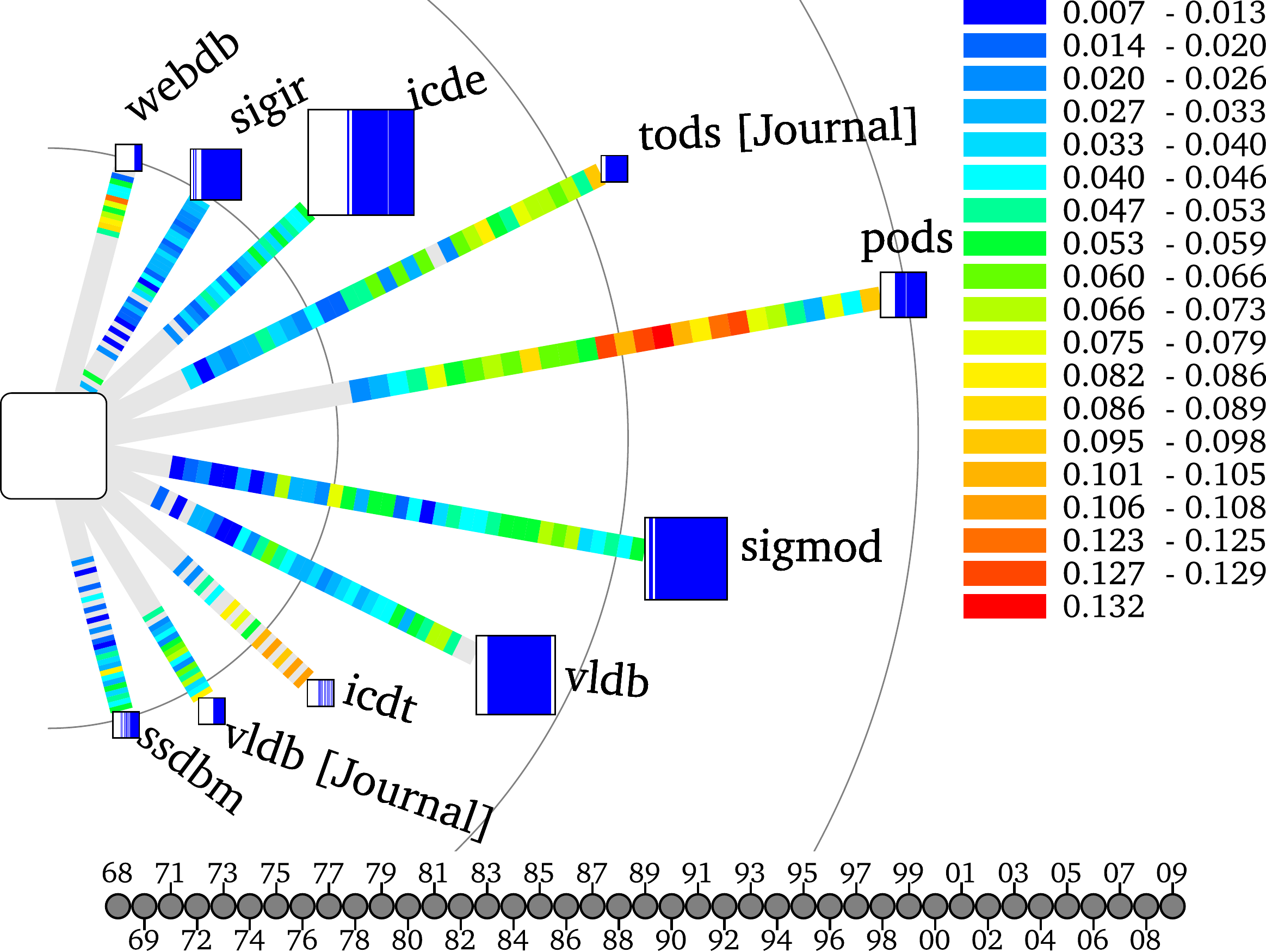}
  \caption{Related streams for the keyword \textit{query}}   \label{img:wordstream}
\end{figure}

Figure \ref{img:streamword} shows which themes were popular at the \textit{Journal of Symbolic Logic} (\textit{JSYML}). \textit{JSYML} is the stream with the most active years in DBLP. We use the words extracted from the publication titles. They are a poor replacement for actual keyword lists but provide acceptable results. Based on the titles from a single year we apply \textit{term frequency - inverse document frequency} (tf-idf)\cite{SaltonB88} to sort out nondescriptive words like \textit{system}. Kuhn and Wattenhofer\cite{KuhnW08} used a similar approach to get thematic descriptions of conferences. The ego node has a time-color filling. Red segments are missing here because \textit{JSYML} was not continued after 2003. We can see different categories of themes. \textit{meeting} and \textit{symbolic} were used from the beginning but do not appear later. \textit{cardinality} was not used at the beginning and \textit{theory} and \textit{logic} appeared at all time. Note that \textit{meeting} and \textit{cardinality} have a similar importance value although they developed differently. Figure \ref{img:wordstream} shows the opposite relation. We see the streams related to the word \textit{query} in the intensive view. tf-idf returns values that are difficult to interpret but the higher the value the stronger the influence. We use presence filling to show when a stream started. For each relevant period we add a segment. If the alter was active in this period the segment would be painted blue otherwise it would be painted white. The \textit{webdb} conference for example was established late and therefore could not use any keywords in early years. We can see that \textit{icdt} is a biennial conference because only every second segment is colored.

There are other relations as well but without additional drawing details. The mot important one is a relation between streams which is rated by comparing communities and themes. We also provide a relation between author and relevant themes.

\subsection{Wikipedia}

As another application we considered relations in the German language Wikipedia. This data set differs from DBLP in size and density of the relation networks. There are 4.6 million authors (including unregistered) and 2.7 million pages. A page is an article, a user page or an administrative page. While an average DBLP author contributed to 2.89 streams, a Wikipedia author modified on average 12.59 articles. This requires more sophisticated caching and pre-calculation strategies for the data interface. There are also differences in the time frame. While DBLP only contains 75 time stamps (1936-2010) the Wikipedia dump we considered was changed on more than 3000 days since 2001. Via the data interface, we define a period as a month and map the respective time stamps to it.

\begin{figure}
 \centering  
  \includegraphics[height=7cm]{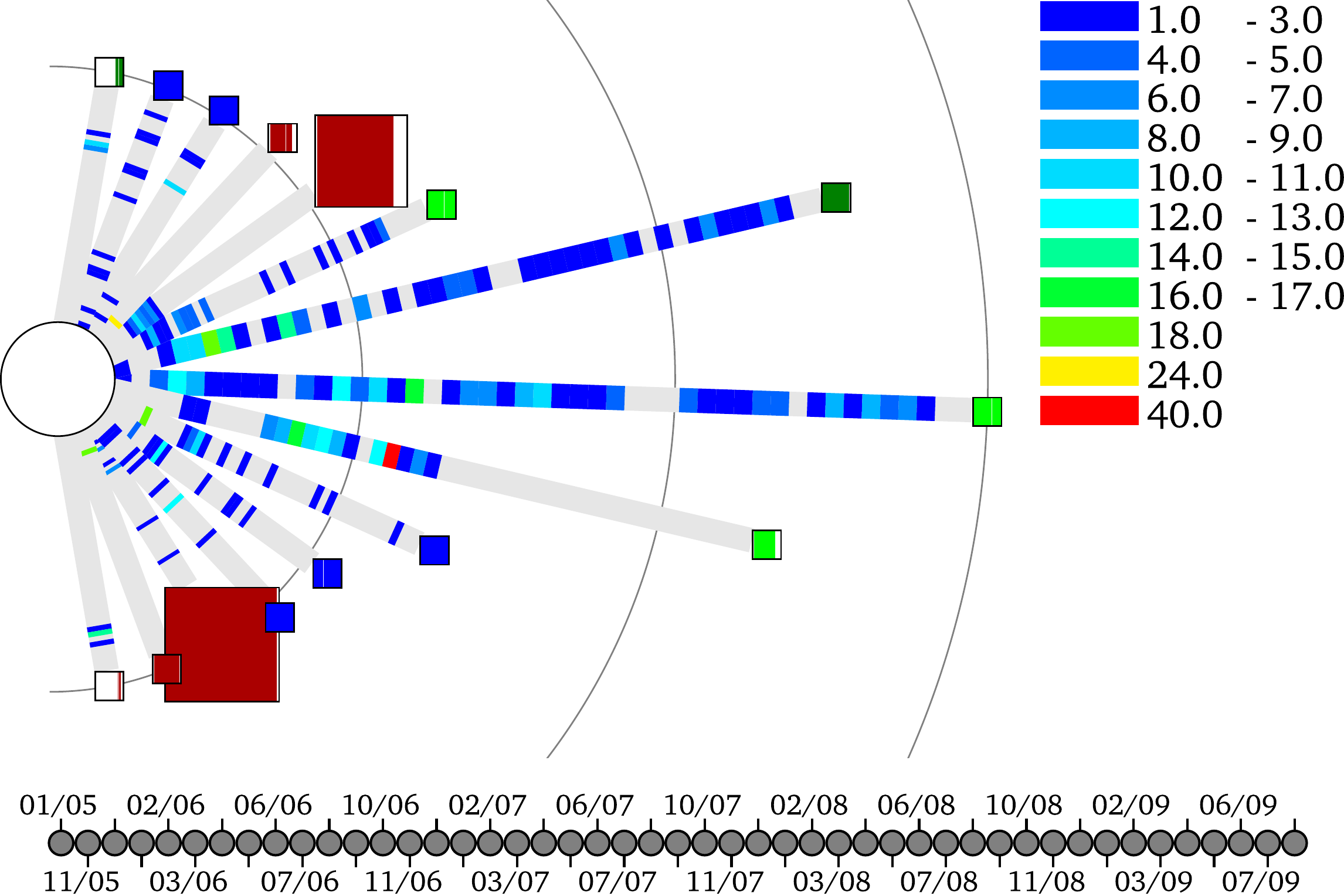}
  \caption{\textit{Gil's} Wikipedia article contributions. Green colors: user pages. Red colors: administration pages, Blue colors: articles.}
  \label{img:wikipedia}
\end{figure}

Figure \ref{img:wikipedia} shows the relation between \textit{Gil} and the articles she modified. \textit{Gil} is administrator so we paint the ego node as a circle. Otherwise, we would have used a rounded rectangle for registered and a triangle for unregistered users. We use presence filling to show in which months an article was changed. The color we use for the presence filling of the alters shows the type of page. 

\section{Evaluation}
\label{sec:evaluation}
The evaluation of our approach consists of a field study in which we observed people using the visual interface and a controlled experiment in the laboratory. We used the DBLP data set (Section \ref{sec:dblp}) in either case.

\subsection{Basic Field Study}
\label{sec:fieldstudy}
In January 2009, we launched a web application using the DBLP data set. For a period of 320 days, we logged which type of graph was requested and which settings were used. To get an approximate mapping of requests and users, we also logged a hash of the session ID. Sessions with a very high number of requests (most probably web bots) were excluded. During the observation, there were 42,068 sessions with a total of 107,683 requests. Most sessions terminated after the first request but 1277 times the user viewed more than ten graphs. The mean length of these long sessions is 30.5 which indicates that the application was actually used for browsing. The time lens was used more often when the session length increased while the use of the search engine dropped. The intensity view was added later so there is no significant data which view was favored.

We also received direct feedback. In Section \ref{sec:timecolor} we already described the remarks on the time-color view and the resulting modifications. The tenor was that the users considered the program to be useful but only after they gathered some experiences with interpreting the drawings. Especially the drawing details seem to pose the risk of confusion and misinterpretations. In general, there were two groups of users. One requested additional information while the other favored simple drawings.

\subsection{Task-based Study}
To get more direct information of the usefulness of our approach we conducted a study in the laboratory.

\noindent \textbf{Participiants: } Two female and eight male undergraduate students aged 22 to 29 participated in the study. All rated themselves as regular computer and web users. Eight participants stated to know the DBLP data set but only one had advanced experience. Nobody was experienced with this framework and the associated visualizations.

\noindent \textbf{Setup: } After a short introduction on the visual interface and DBLP, we gave the participants time to familiarize themselves with the application. Then we asked them to complete three groups of tasks ($G_1$, $G_2$ and $G_3$) with three problems each.

\begin{enumerate}
 \item[$G_1$] Tasks could be solved by analyzing a single graph.
 \item[$G_2$] Tasks required using a specific feature (for example time lens) or a specific view. 
 \item[$G_3$] \textit{Open tasks}. We asked participants to explore the neighborhood of a given entity. Noteworthy constellations should be reported.
\end{enumerate}

Because no participiant was experienced in the application domain, we had to limit the complexity of the tasks in $G_3$. For $G_1$ and $G_2$ we expected short answers to specific questions. For $G_3$ we were mainly interested in how the users applied their experience from $G_1$ and $G_2$. We randomized the order of tasks within each group to counter learning effects. We observed the participants and logged which graphs they requested and which parameter they used. The tasks had to be completed in 30 minutes. At the end of the session, the participants were asked to answer open end questions on what they liked and didn't like on the framework.

\noindent \textbf{Task Results: } 
Eight participants solved all $G_1$ tasks  and six all $G_2$ tasks. Nobody failed more than one task in a group. For each task we asked if completing it had been easy, slightly difficult or difficult. In 55 out of 60 cases, the task was rated easy. The remaining cases were considered to be slightly difficult. Eight participants were able to find interesting constellations in $G_3$. The others reported only uninteresting details or actual misinterpretations. Nine participants stated that the lack of knowledge on the DBLP data set was the major problem. All were positive to find more if given additional time. By observing the users and analyzing the log files we found out that:

\begin{itemize}
\item All Tasks (or parts of task) which required to tell the relevance of an entity were correctly solved.
\item At first, the users preferred the time-color view. A task in $G_2$ forced them to use the intensity view. After that, this view was preferred.
\item In $G_2$ half of the users browsed to a new graph using the search function rather than the connection menus. With growing experience this
behavior changed. In $G_3$ all participants used the connection menus if possible. 
\item Not understanding drawing details was the major cause of errors. While there was no problem with the fraction filling, many users misinterpreted
the fillings of the stream nodes (see Figure \ref{img:personstream}). Eight participiants reported that they were confused because the filling used colors 
which did not appear in the bottom bar.
\item Tooltips were used more often than we expected. All participants tried to validate their interpretation of the drawing with textual information
if possible.
\item The time lens was considered an important part of the application. It was used in many cases, even if it did not contribute to the solution.
\end{itemize}

In general, the comments were positive. The users liked the small and clear graph drawings and the idea of presenting a rating by the length of an edge. Only one person criticized the fact that \textit{close} related alters are placed \textit{far} away from the ego. Like in the field study feedback, learning to use the application was a major issue. Four participants explicitly stated that they had to learn how to use the application first. There also was a significant learning effect. Unlike the feedback from the field study, the participants requested more textual information. Four persons explicitly mentioned text integrated in the drawings. For example, the relevance value should be visible next to the nodes. Two proposed an additional view which should contain the information in tabular form. This supports our finding that tooltips are frequently used. The tooltips were consistantly mentioned as a positive aspect.

Four participants stated that the information density was too high. But, like in the field study feedback, all users posted ideas on what information should be added. Three participants proposed additional drawing details like patterns or more types of node shapes. This shows that a graph definition has to be done very carefully with respect to the user group. The definer must not give in on the wish to integrate as much information as possible but has to make reasonable selections. 

\section{Related Work}
Aggregated approaches show the temporal data in a single drawing. LifeLines\cite{PlaisantMRWS96} visualizes a person's disease pattern. For each condition there is a timeline, i.e., a horizontal bar along a time axis. Coloration and thickness of these bars change to show the status of the condition at different times. TimeRadarTrees\cite{BurchD08} uses a radial drawing to show how several entities are related with each other. Unlike our approach, there is no ego and all relations between the visible nodes are displayed. Instead of node-link, it uses colored segments which fill a circle with multiple layers. The approach is limited to small graphs but supports hierarchical nesting to compensate this. Segments at the perimeter represent recent events while those near the center represent old influences. The intensity view is similar to this drawing but has a much lower information density. ConfSearch\cite{KuhnW08} searches DBLP for relations between conferences, authors and keywords. The related entities of an ego are presented as a rated list with additional information. ConfSearch does not show the evolution of relations, but we adopted some of the rating functions used for the examples in Section \ref{sec:dblp}.

Many systems combine different types of visualizations. Paper Lense \cite{LeeCRB05} and Facet Lense \cite{LeeSRCT09} provide bar charts, textual result lists and nested node drawings to show entities in faceted data sets. Among others, data can be plotted against time, like the number of an author's publications by year. Both tools provide extensive filtering and sorting functions. The DB-Browser\cite{WeberRWLK06} features similar views including simple graph drawings. PaperLense and DB-Browser visualize DBLP data. Both provide aggregated information like the number of joint papers for two given authors.

Not all approaches use the entity and relation abstraction. The ThemeRiver\cite{HavreHWN02} application shows how the frequency of a term in a set of documents changes over time. The results for multiple terms are presented as a plot where one axis shows the time and the other axis the frequency. 

\section{Conclusion And Future Work}
In this paper, we presented a framework that can visualize relations in any given data set. Large and complex relation networks are reduced to small graphs. 
The drawings of these graphs resemble ego-centered networks and convey information on when and how strong a relation evolved. The drawings are part of a visual interface which supports the user in understanding the data and links the graphs with each other. Experiments from an online application and the results of a basic usability study indicated that our approach is useful though it poses the risk of generating graphs which overstrain the user. Future work will have to address the problem of how difficult it is to understand the drawing details. These studies will also have to include other drawing details.

There is no clear definition of an ego-centered graph in the literature. Many definitions allow additional edges between the alters or even nodes that are in no direct connection with the ego. The decision to include a node or an edge is usually based on complex strategies or the intuition of the person who generates the graph. It is unclear whether the inverted ego network could be extended this way and whether this would benefit the user.

\bibliographystyle{abbrv}
\bibliography{vis}{}

\end{document}